\documentclass[10pt,letterpaper]{article}
\usepackage[top=0.85in,left=2.75in,footskip=0.75in]{geometry}
\usepackage{changepage}
\usepackage[utf8]{inputenc}
\usepackage{textcomp,marvosym}
\usepackage{fixltx2e}
\usepackage{amsmath,amssymb}
\usepackage{cite}
\usepackage{nameref,hyperref}
\usepackage{microtype}
\DisableLigatures[f]{encoding = *, family = * }
\usepackage{rotating}
\usepackage{bbold}
\usepackage{dsfont}

\usepackage{amsmath}
\DeclareMathOperator{\sign}{sign}

\usepackage{mathrsfs}

\usepackage{siunitx}

\usepackage{outlines}
\usepackage{enumitem}
\setlist[enumerate,2]{label=\roman*)}
\setlist[enumerate,3]{label=\alph*)}

\newenvironment{myenumerate}
{ \begin{enumerate}
    \setlength{\itemsep}{0pt}
    \setlength{\parskip}{0pt}
    \setlength{\parsep}{0pt}     }
{ \end{enumerate}                  } 

\DeclareMathOperator*{\argmax}{arg\,max}

\raggedright
\usepackage[aboveskip=1pt,labelfont=bf,labelsep=period,justification=raggedright,singlelinecheck=off]{caption}

\makeatletter
\renewcommand{\@biblabel}[1]{\quad#1.}
\makeatother

\date{}

\usepackage{lastpage,fancyhdr,graphicx}
\fancyheadoffset[L]{2.25in}
\fancyfootoffset[L]{2.25in}

\begin{document}

\begin{flushleft}
{\Large
\textbf\newline{Cell cycle time series gene expression data encoded as cyclic attractors in Hopfield systems}
}
\newline
\\
Anthony Szedlak\textsuperscript{1},
Spencer Sims\textsuperscript{1},
Nicholas Smith\textsuperscript{2},
Giovanni Paternostro\textsuperscript{3},
Carlo Piermarocchi\textsuperscript{1,*},
\\
\bf{1} Department of Physics and Astronomy, Michigan State University, East Lansing, MI, USA
\\
\bf{2} Salgomed Inc., Del Mar, CA, USA
\\
\bf{3} Sanford Burnham Prebys Medical Discovery Institute, La Jolla, CA, USA
\\

%
%





* E-mail: Corresponding carlo@pa.msu.edu
\end{flushleft}
\section*{Abstract} 

Modern time series gene expression and other omics data sets have enabled unprecedented resolution of the dynamics of cellular processes such as cell cycle and response to pharmaceutical compounds. In anticipation of the proliferation of time series data sets in the near future, we use the Hopfield model, a recurrent neural network based on spin glasses, to model the dynamics of cell cycle in HeLa (human cervical cancer) and \textit{S. cerevisiae} cells. We study some of the rich dynamical properties of these cyclic Hopfield systems, including the ability of populations of simulated cells to recreate experimental expression data and the effects of noise on the dynamics. Next, we use a genetic algorithm to identify sets of genes which, when selectively inhibited by local external fields representing gene silencing compounds such as kinase inhibitors, disrupt the encoded cell cycle. We find, for example, that inhibiting the set of four kinases \textit{BRD4}, \textit{MAPK1}, \textit{NEK7}, and \textit{YES1} in HeLa cells causes simulated cells to accumulate in the M phase. Finally, we suggest possible improvements and extensions to our model.

\section*{Author Summary} 

Cell cycle -- the process in which a parent cell replicates its DNA and divides into two daughter cells -- is an upregulated process in many forms of cancer. Identifying gene inhibition targets to regulate cell cycle is important to the development of effective therapies. Although modern high throughput techniques offer unprecedented resolution of the molecular details of biological processes like cell cycle, analyzing the vast quantities of the resulting experimental data and extracting actionable information remains a formidable task. Here, we create a dynamical model of the process of cell cycle using the Hopfield model (a type of recurrent neural network) and gene expression data from human cervical cancer cells and yeast cells. We find that the model recreates the oscillations observed in experimental data. Tuning the level of noise (representing the inherent randomness in gene expression and regulation) to the ``edge of chaos'' is crucial for the proper behavior of the system. We then use this model to identify potential gene targets for disrupting the process of cell cycle. This method could be applied to other time series data sets and used to predict the effects of untested targeted perturbations.






\section*{Introduction}

Originally proposed by Conrad Waddington in the 1950s~\cite{waddington2014strategy} and Stuart Kauffman in the 1970s~\cite{kauffman1971differentiation}, analysis of biological processes such as cellular differentiation and cancer development using attractor models -- dynamical systems whose configurations tend to evolve toward particular sets of states -- has gained significant traction over the past decade~\cite{huang2005cell,alvarez2008floral,huang2009cancer,demicheli2010gene,choi2012attractor,yuan2016molecular,pusuluri2017exploring,udyavar2017novel,wooten2017mathematical,yuan2017cancer}. One such attractor model, the Hopfield model~\cite{hopfield1982neural}, is a type of recurrent artificial neural network based on spin glasses. It was designed with the ability to recall a host of memorized patterns from noisy or partial input information by mapping data directly to attractor states. A great deal of analytical and numerical work has been devoted to understanding the statistical properties of the Hopfield model, including its storage capacity~\cite{peretto1988learning}, correlated patterns~\cite{kanter1987associative}, spurious attractors~\cite{amit1985storing}, asymmetric connections~\cite{derrida1987exactly}, embedded cycles~\cite{nishimori1990retrieval}, and complex transition landscapes~\cite{personnaz1986collective}. Due to its prescriptive, data-driven design, the Hopfield model has been applied in a variety of fields including image recognition~\cite{nasrabadi1991object,cote1997hopfield} and the clustering of gene expression data~\cite{maetschke2014characterizing}. It has also been used to directly model the dynamics of cellular differentiation and stem cell reprogramming~\cite{lang2014epigenetic}, as well as targeted inhibition of genes in cancer gene regulatory networks~\cite{szedlak2014control}.

Techniques for measuring large scale omics data, particularly transcriptomic data from microarrays and RNA sequencing (RNA-seq), have become standard, indispensable tools for measuring the states of complex biological systems~\cite{wang2009rna,marguerat2010rna,trapnell2012differential}. However, analysis of the sheer variety and vast quantities of data these techniques produce requires the development of new mathematical tools. 
Inference and topological analysis of gene regulatory networks has garnered much attention as a method for distilling meaningful information from large datasets~\cite{margolin2006aracne,hecker2009gene,lee2011prioritizing,marbach2012wisdom,cahan2014cellnet,ong2015scalable,szedlak2016evolutionary}. But because life is a non-equilibrium phenomenon that can only be truly understood at the dynamical level, there is a growing need to develop new methods for analyzing time series data.
As experimental methods continue to improve, more and more high-resolution time series omics and even multi-omics~\cite{chen2012personal} data sets will inevitably become available. Here, we demonstrate that time series omics data (in this case, transcriptomic data) representing cyclic biological processes can be encoded in Hopfield systems, providing a new model for analyzing the dynamics of, and exploring effects of perturbations to, such systems. 

The dynamics of cell cycle (CC) -- the process in which a parent cell replicates its DNA and divides into two daughter cells -- is both scientifically interesting and therapeutically important. Even relatively simple simulated systems such as an isolated, positively self-regulating gene subject to noise can exhibit rich dynamical behavior~\cite{assaf2013extrinsic}; but like many biological processes, the proper functioning of CC requires the decentralized, coordinated action of hundreds of genes. CC thus provides researchers with a convenient case study of self-organization in a noisy environment. CC is also an upregulated process in many forms of cancer~\cite{evan2001proliferation,malumbres2009cell,hanahan2011hallmarks,diaz2013targeting}, and control of CC using pharmaceutical compounds such as kinase inhibitors is a critical goal in cancer research. The combinatorics of selectively inhibiting sets of genes makes exhaustive experimental searches difficult or impractical~\cite{feala2010systems}. However, network-based mathematical models such as the one presented here enable researchers to examine the effects of perturbations to complex systems~\cite{csermely2005efficiency,agoston2005multiple} by testing potential inhibition targets \textit{in silico}. The efficacy of these predictions can then be experimentally validated or invalidated, providing new information and insights to further refine models.

The remainder of this article is structured as follows. In the Models section we first discuss how periodic genes were identified in the time series gene expression data sets, and how Boolean attractors were extracted from the continuous data (explained in greater detail in the Methods section). We then introduce the Hopfield model and discuss the specific form of the coupling matrix used in this application. We discuss how to interpret the results of Hopfield simulations in the context of gene expression and cells. We also explain the objective function used by the genetic algorithm to identify potential inhibition targets, designed with the intention of disrupting CC. In the Results and Discussion section, we show that this model qualitatively recreates experimental gene expression data, and we demonstrate and analyze some dynamical properties of the delayed Hopfield model, including the role played by noise. We include supplementary videos to emphasize the dynamical nature of this model. Optimal control fields for both unconstrained searches (in which any gene may be inhibited) and searches constrained to kinases in HeLa cells are discussed. Finally, we recap our results and suggest possible improvements and generalizations to our methods in the Conclusions section.

\section*{Models} 

\subsection*{Periodic gene selection}

\begin{figure}
\begin{center}
\includegraphics[width=1.0\textwidth]{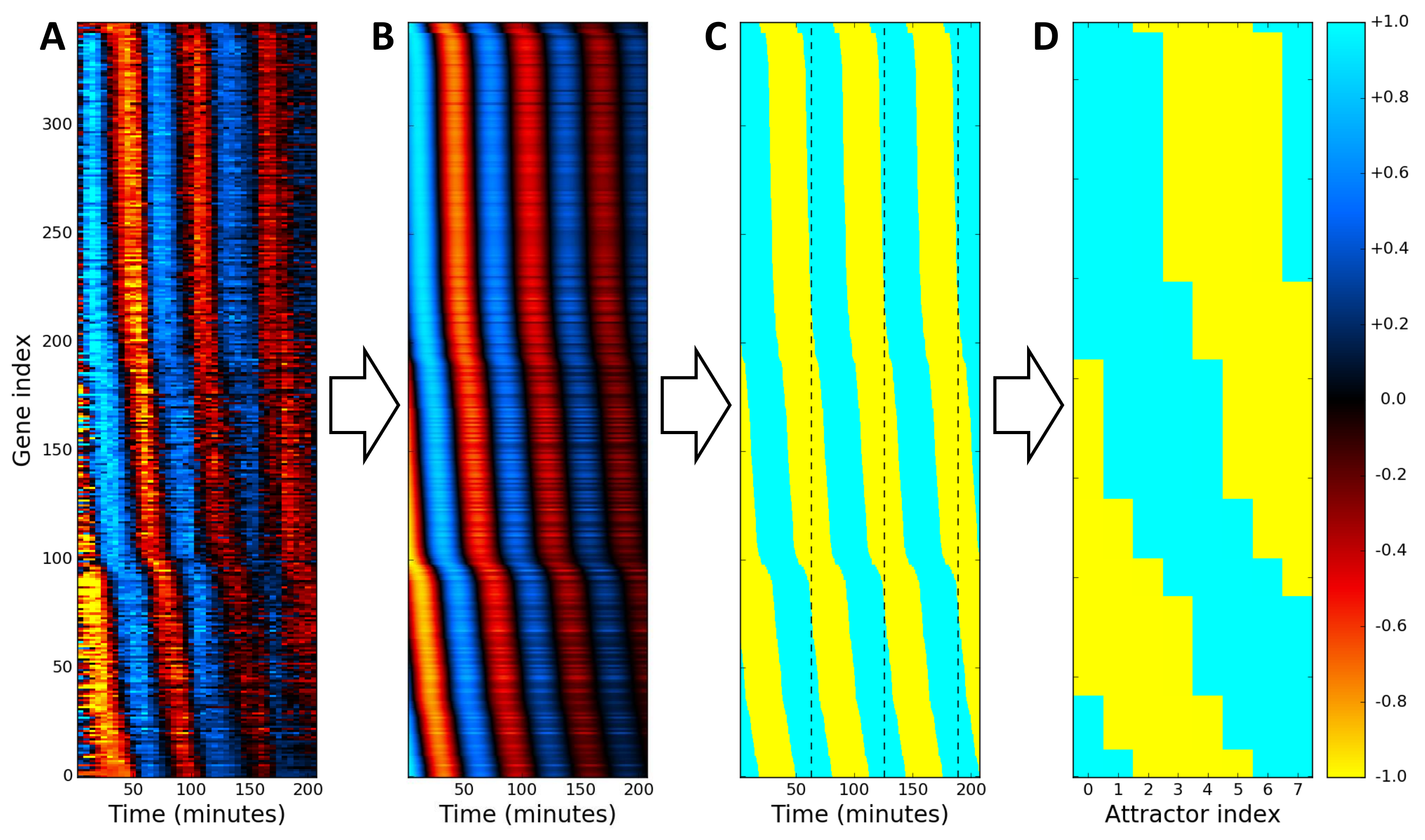}
\end{center}
\caption{{\bf Obtaining attractors from expression data.} (A) Heat map of the expression of all detected periodic genes from~\cite{eser2014periodic} sorted by their fitted phases. (B) Fitted gene expression. (C) Boolean form of fitted expression, separated into periods by dashed black lines. (D) Final set of $p=8$ attractors taken from one period.}
\label{fig:snakePlot}
\end{figure}

Microarray and RNA-seq time series data sets were obtained from Eser et al.~(\textit{S. cerevisiae})~\cite{eser2014periodic} and Dominguez et al.~(HeLa, human cervical cancer)~\cite{dominguez2016high}. For consistency and due to its higher resolution, the \textit{S. cerevisiae} data set was chosen to produce all images and movies in this article, but both data sets were analyzed. In order to encode these CC data sets into the Hopfield model, periodic genes needed to be identified, their frequencies and phases computed, and their expression converted from continuous to Boolean form. As detailed in the Methods section, decaying sinusoids were fitted to the trajectory of each gene $i$, and genes with sufficiently high quality fits were kept. This resulted in 379 periodic genes in \textit{S. cerevisiae} and 519 periodic genes in HeLa cells. Figure~\ref{fig:snakePlot}A shows a heat map of the expression of all periodic genes detected in the Eser data set sorted by their fitted phases, and Figure~\ref{fig:snakePlot}B shows the same genes with the fitted expression curves. These fitted curves were converted from continuous values $x_i(t)\geq0$ to Boolean values $\xi_i(t)=\pm1$ (over/underexpressed) as shown in Figure~\ref{fig:snakePlot}C. Finally, one CC period was divided into eight uniformly spaced states $\{\xi_i^\mu\} = \{\xi_i^0,\xi_i^1,\ldots,\xi_i^7\}$ with $\xi_i^\mu=\pm1$. These states, shown in Figure~\ref{fig:snakePlot}D, were used as the eight attractor patterns in the Hopfield model.

\subsection*{The Hopfield model}

The Hopfield model~\cite{hopfield1982neural} is an Ising model whose configuration is defined by $N$ spins $\sigma_i(t)$ at integer time $t$. The state of each node (gene) takes one of two values, $\sigma_i(t)=\pm1$ (over/underexpressed). The coupling matrix $J_{ij}$ defines the strength and sign of the signal sent from node $j$ to node $i$, and its construction is discussed in the following subsection. The total field at node $i$ at time $t$ is given by
\begin{equation}
h_i(t) = \sum_{j}J_{ij}\sigma_j(t) + h_i^\text{ext} \text{ ,}
\end{equation}

\noindent where $\sum_{j}J_{ij}\sigma_j(t)$ is the internal field at node $i$ due to its coupling with all nodes $j$ and $h_i^\text{ext}$ is an optional external field applied to node $i$ representing the action of therapeutic compounds, e.g.~kinase inhibitors. The dynamical update rule is given by
\begin{equation}
\sigma_i(t+1) =
\begin{cases}
    +1              & \text{with probability } (1+e^{-2h_i(t)/T})^{-1}\\
    -1              & \text{otherwise}
\end{cases} \text{ ,}
\label{eq:hopfieldUpdateRule}
\end{equation}

\noindent where the factor of 2 in the exponent is conventional and $T$ is an effective temperature representing the level of noise (not a physical temperature). Biologically, this noise represents the effects of all kinds of biochemical fluctuations present in cells. Note that for $h_i(t)\rightarrow\pm\infty$, $\sigma_i(t+1)=\pm1$; for $T\rightarrow\infty$, $\sigma_i(t+1)=\pm1$ with equal probability; and for $T\rightarrow0$, $\sigma_i(t+1) = \text{sign}\left(h_i(t)\right)$.

The update rule from Eq.~\ref{eq:hopfieldUpdateRule} may be implemented in various ways. The synchronous scheme updates the state of all nodes in the system at every time step, but this is sensible only if the simulated system has a central pacemaker coordinating the activity of all nodes. A more appropriate choice for decentralized systems like gene regulatory networks is the asynchronous scheme in which the state of a randomly chosen subset of nodes is updated at each time step. Here, we use the asynchronous scheme with update probability $0.2$ for each node.

\subsection*{Coupling matrix}

In the canonical Hopfield model, the coupling matrix is constructed to store a set of $p$ linearly independent (i.e.~distinct) Boolean patterns $\xi_i^\mu=\pm1$ as point attractors, where $i=0,1,\ldots,N-1$ is the node index and $\mu=0,1,\ldots,p-1$ is the pattern index. The point attractor coupling matrix $J_{ij}^\prime$ is given by
\begin{equation}
J_{ij}^\prime = \frac{1}{N} \sum_{\mu\nu} \xi_i^\mu \left({Q^{-1}}\right)_{\mu\nu} \xi_j^\nu \text{ ,}
\label{eq:pointCouplingMatrix}
\end{equation}

\noindent where~\cite{kohonen1974adaptive,kanter1987associative}
\begin{equation}
Q_{\mu\nu} = \frac{1}{N} \sum_{i} \xi_i^\mu \xi_i^\nu \text{ .}
\end{equation}

\noindent With this coupling matrix and $T=0$, if at some time $t$ the configuration is given by $\sigma_i(t)=\xi_i^\mu+\delta_i$ for a small perturbation $\delta_i$, then $\lim_{t\rightarrow\infty}\sigma_i(t)=\xi_i^\mu$. Note that this formulation means $\pm\xi_i^\mu$ are both attractors of the system.

A simple modification~\cite{personnaz1986collective} to Eq.~\ref{eq:pointCouplingMatrix} produces a cyclic attractor coupling matrix $\widetilde{J}_{ij}$, constructed according to
\begin{equation}
\widetilde{J}_{ij} = \frac{1}{N} \sum_{\mu\nu} \xi_i^{\text{mod}_p \left( \mu+1 \right)} \left({Q^{-1}}\right)_{\mu\nu} \xi_j^\nu \text{ .}
\label{eq:cyclicCouplingMatrix}
\end{equation}

\noindent At $T=0$, this coupling matrix cyclically maps through the sequence of $p$ patterns
\begin{equation}
\xi_i^0\rightarrow
\xi_i^1\rightarrow\ldots\rightarrow
\xi_i^{p-2}\rightarrow
\xi_i^{p-1}\rightarrow
\xi_i^0\rightarrow
\xi_i^1\rightarrow
\ldots
\end{equation}

\noindent or their negatives. For the remainder of this article, all attractor indexing is understood to be modulo $p$.

A delayed cyclic Hopfield model may be constructed by combining the point and cyclic attractor matrices into one coupling matrix,
\begin{equation}
J_{ij}(\lambda) = (1-\lambda)J^\prime_{ij}+\lambda\widetilde{J}_{ij} \text{ ,}
\end{equation}

\noindent for an adjustable transition strength parameter $\lambda$ with $0\leq\lambda\leq1$. If $\sigma(t)=\xi_i^\mu$, $\lambda\ll1$, and $T=0$, the point attractor term dominates and $\sigma_i(t)=\sigma(t+\Delta t)$ for all $\Delta t=1,2,\ldots$. If $T>0$, however, stochastic fluctuations eventually push the configuration out of the basin of attraction of the $\mu^\text{th}$ attractor and into the $(\mu+1)^\text{th}$ basin, then eventually to the $(\mu+2)^\text{th}$ basin, and so on. The dynamics of the delayed cyclic Hopfield model are thus governed by noise-induced transitions.

Due to the sinusoidal nature of the gene expression in these CC data sets, however, the attractors are structured such that $\xi_i^\mu = -\xi_i^{\mu+4}$, making $Q_{\mu\nu}$ rank deficient and thus noninvertible. Because the definition of $J_{ij}$ automatically guarantees that if any sequence $\{+\xi_i^\mu\}$ is an attractor, then $\{-\xi_i^\mu\}$ is also an attractor, encoding the sequence
\begin{equation}
\xi_i^0\rightarrow\xi_i^1\rightarrow\xi_i^2\rightarrow\xi_i^3\rightarrow\xi_i^4 \left(=-\xi_i^0\right)
\end{equation}

\noindent automatically encodes the sequence
\begin{equation}
\xi_i^4\rightarrow\xi_i^5\rightarrow\xi_i^6\rightarrow\xi_i^7\rightarrow\xi_i^0 \left(=-\xi_i^4\right) \text{ .}
\end{equation}

\noindent In this special case of sinusoidal trajectories, the limits of summation in Eqs.~\ref{eq:pointCouplingMatrix}-\ref{eq:cyclicCouplingMatrix} need only run over the first four indices, $\mu=0,1,2,3$.

Finally, to reflect the fact that real gene regulatory networks are sparse, weak edges were removed by setting all elements of the coupling matrix with $|J_{ij}|<\text{median}(|J|)$ to zero, where $|J|$ is element-wise absolute value.

\subsection*{Biological interpretation of the dynamics}

Extracting biological meaning from this model requires defining some convenient coarse-grained quantities. The overlap of the state vector $\sigma_i(t)$ with the $\mu^\text{th}$ pattern is given by
\begin{equation}
m^\mu(t) = \frac{1}{N}\sum_{i} \sigma_i(t) \xi_i^\mu \text{ ,}
\end{equation}

\noindent where $-1\leq m^\mu(t)\leq+1$. The overlap measures the similarity between the (discretized) experimental and simulated gene expression profiles, and $m^\mu(t)=+1$ means there is perfect agreement between the simulated cell's expression and pattern $\mu$.

A single configuration vector $\sigma_i(t)$ represents the expression profile of a single cell. For many cells $\kappa$, let $\sigma_{ik}(t)$ be the expression of gene $i$ in cell $k$. Define
\begin{equation}
m^\mu_k(t) = \frac{1}{N}\sum_i \sigma_{ik}(t) \xi_i^\mu
\label{eq:ensembleOverlap}
\end{equation}

\noindent as the overlap of cell $k$ with attractor $\mu$. Because the microarray and RNA-seq data used here report the gene expression averaged over many cells, it is appropriate to define the population-averaged (i.e.~ensemble-averaged) expression,
\begin{equation}
\left\langle \sigma_i(t) \right\rangle_\mathcal{K} = \frac{1}{\kappa} \sum_{k=0}^{\kappa-1} \sigma_{ik}(t) \text{ ,}
\label{eq:avgGeneExpression}
\end{equation}

\noindent which has $-1\leq\left\langle \sigma_i(t) \right\rangle_\mathcal{K}\leq+1$.

Rather than work with a continuous vector quantity like $m^\mu_k(t)$, each cell can simply be identified as being in a discrete phenotypic state at any given time. Define the state of cell $k$ as
\begin{equation}
s_k(t) = \argmax_\mu m^\mu_k(t) \text{ ,}
\label{eq:cell_state}
\end{equation}

\noindent i.e.~the index of the attractor with maximum overlap, which may be interpreted as cell $k$'s phenotype. To better understand population-level dynamics, define the discrete probability distribution $P_\mu(t)$ as the fraction of $\kappa$ cells with $s_k(t)=\mu$; that is, $P_\mu(t)$ is the probability that a randomly chosen cell is in state $\mu$ at time $t$. Finally, define the time-averaged distribution of states as
\begin{equation}
\left\langle P_\mu \right\rangle_\mathscr{T} = \frac{1}{\tau} \sum_{t=0}^{\tau-1} P_\mu(t)
\end{equation}

\noindent for a window of time $\tau$.

For each data set, $T$ and $\lambda$ were tuned to the ``edge of chaos''~\cite{kauffman1991coevolution} such that the cyclic attractor was preserved and the time between transitions was approximately constant, but the system was sensitive enough to perturbations that some targeted inhibitions produced noticeable changes in $\left\langle P_\mu \right\rangle_\mathscr{T}$. See~\nameref{S1_Table} for a list of parameters used for each data set.

\subsection*{Gene inhibition optimization}

In this application, the goal is to identify perturbations that halt or retard the encoded cyclic attractor. A standard genetic algorithm (GA; explained in~\nameref{sup:genetic_alg}) was employed to identify an optimal control field $h^\text{opt}_i$ that maximized a given objective function $f(h^\text{ext}_i)$,
\begin{equation}
h^\text{opt}_i = \argmax_{h^\text{ext}_i}{f\left(h^\text{ext}_i\right)} \text{ ,}
\end{equation}

\noindent where $h^\text{ext}_i$ is the control vector given by
\begin{equation}
h^\text{ext}_i =
\begin{cases}
    -\infty        & \text{if gene $i$ is targeted} \\
    0              & \text{otherwise}
\end{cases}
\end{equation}

\noindent for a fixed number of targets (nonzero elements) $n_\text{targ}$. Only negative control fields are used here to simulate the effects of targeted gene inhibition from pharmaceutical compounds. The objective function used here is $\left\langle P_\mu \right\rangle_\mathscr{T}$, meaning that the optimal control field maximizes the time-averaged number of cells occupying a particular attractor state $\mu$. This search was conducted across all attractors $\mu$ to determine the controllability of each attractor state.


\section*{Results and Discussion} 

{\bf\textit{Note:}} The supporting information can be downloaded from this temporary Dropbox link: \url{http://bit.ly/2tEekWK}. For convenience, the four supplementary videos can also be viewed using the following unlisted YouTube links:
\begin{myenumerate}
\item S1 Video: \url{https://youtu.be/LOUjRftAeYM}
\item S2 Video: \url{https://youtu.be/pfGbla_LeGI}
\item S3 Video: \url{https://youtu.be/pfatwL7TusQ}
\item S4 Video: \url{https://youtu.be/RT7uNAGDcyA}
\end{myenumerate}

\subsection*{Dynamical behavior}

\begin{figure}
\begin{center}
\includegraphics[width=1.0\textwidth]{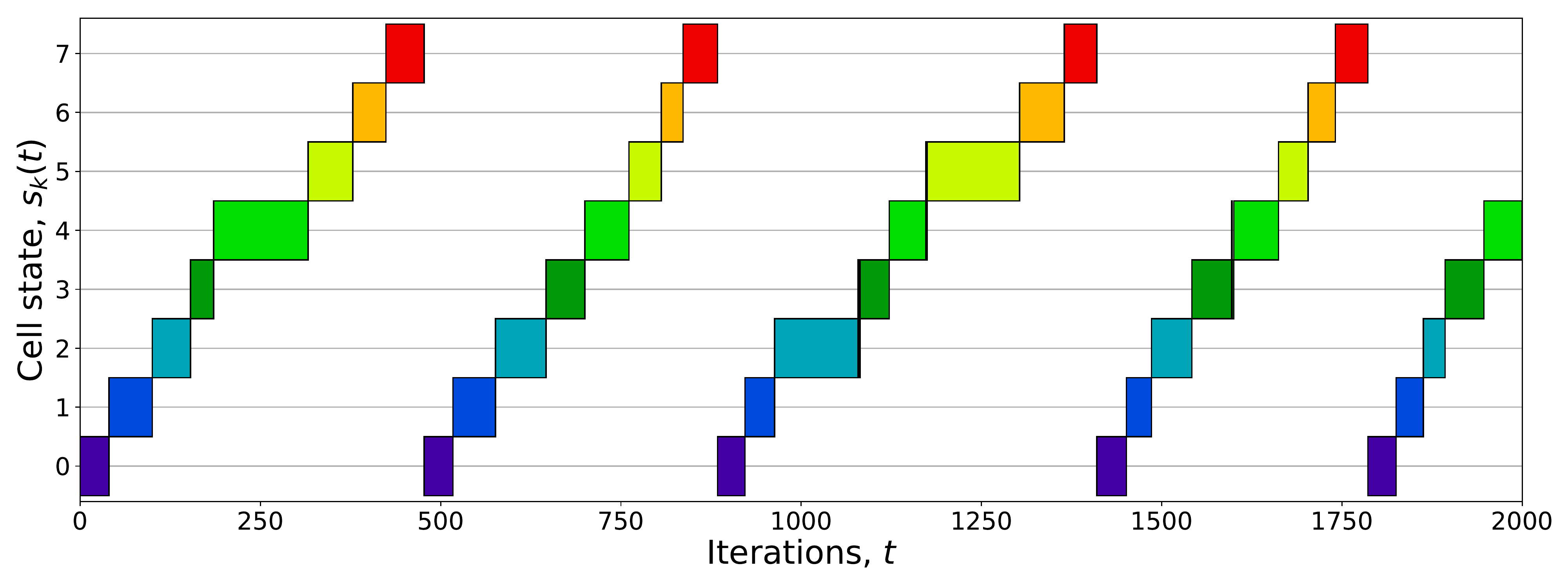}
\end{center}
\caption{{\bf Unperturbed cell state versus time.} Boxes indicate $s_k(t)$, i.e.~the index of the attractor with maximum overlap at time $t$. The system began with the configuration $\sigma_i(0)=\xi_i^0$ and was allowed to evolve according to the Hopfield signaling rules with zero external field, mapping cyclically through the set of eight attractors. The pattern and cycle durations vary due to the system's stochasticity.}
\label{fig:eserUnperturbed}
\end{figure}

Figure~\ref{fig:eserUnperturbed} shows the time evolution of $s_k(t)$ for a single simulated cell using the attractors derived from~\cite{eser2014periodic}. As expected, the system progresses cyclically through the eight attractor states. The duration of each cycle varies somewhat due to the stochasticity in the update rule from Eq.~\ref{eq:hopfieldUpdateRule}.

\begin{figure}
\begin{center}
\includegraphics[width=1.0\textwidth]{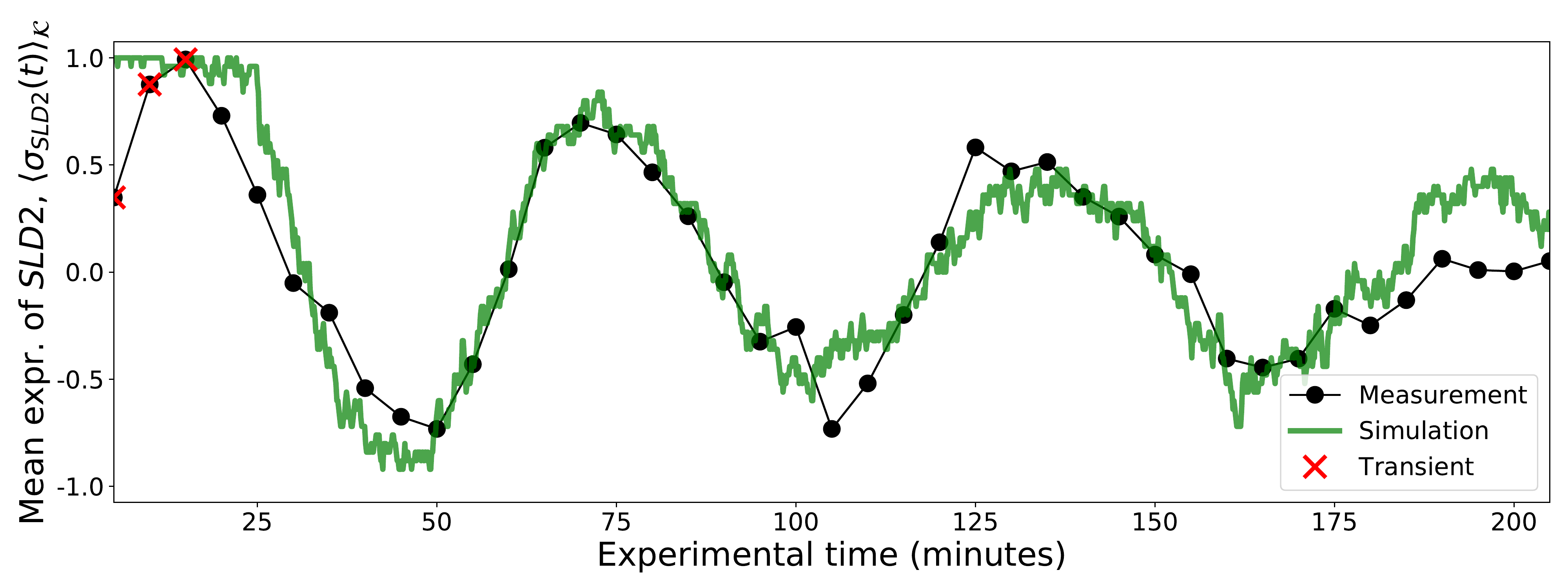}
\end{center}
\caption{{\bf Measured and simulated gene expression.} The measured expression of the \textit{S. cerevisiae} gene \textit{SLD2} from~\cite{eser2014periodic} was scaled to the range $[-1,+1]$ and is shown as a black beaded curve, and the population-averaged expression of the same gene as defined in Eq.~\ref{eq:avgGeneExpression} for $\kappa=50$ cells is shown in green, with the $t$ axis rescaled by eye to match experimental time. Transient points (red X's) were ignored when fitting Eq.~\ref{eq:firstFit}.}
\label{fig:exampleGene}
\end{figure}

Although the gene expression for each simulated cell $k$ has $\sigma_{ik}(t)=\pm1$, the population-averaged expression has $-1\leq\langle\sigma_i(t)\rangle_\mathcal{K}\leq+1$, and for many cells initially synchronized with $\sigma_{ik}(0)=\xi_i^0$ for all $k$, $\langle\sigma_i(t)\rangle_\mathcal{K}$ successfully recovers the experimentally observed decaying sinusoidal gene expression. Figure~\ref{fig:exampleGene} shows a comparison between the experimental expression $x_i(t)$ from the Eser data set and the mean simulated expression $\left\langle \sigma_i(t) \right\rangle_\mathcal{K}$ with $\kappa=50$ for $i=\text{\textit{SLD2}}$, one of the genes responsible for initiating DNA replication in \textit{S. cerevisiae}~\cite{muramatsu2010cdk,bruck2014replication}. The simulation time $t$ was rescaled by eye to align the simulated and experimental curves.

\begin{figure}
\begin{center}
\includegraphics[width=1.0\textwidth]{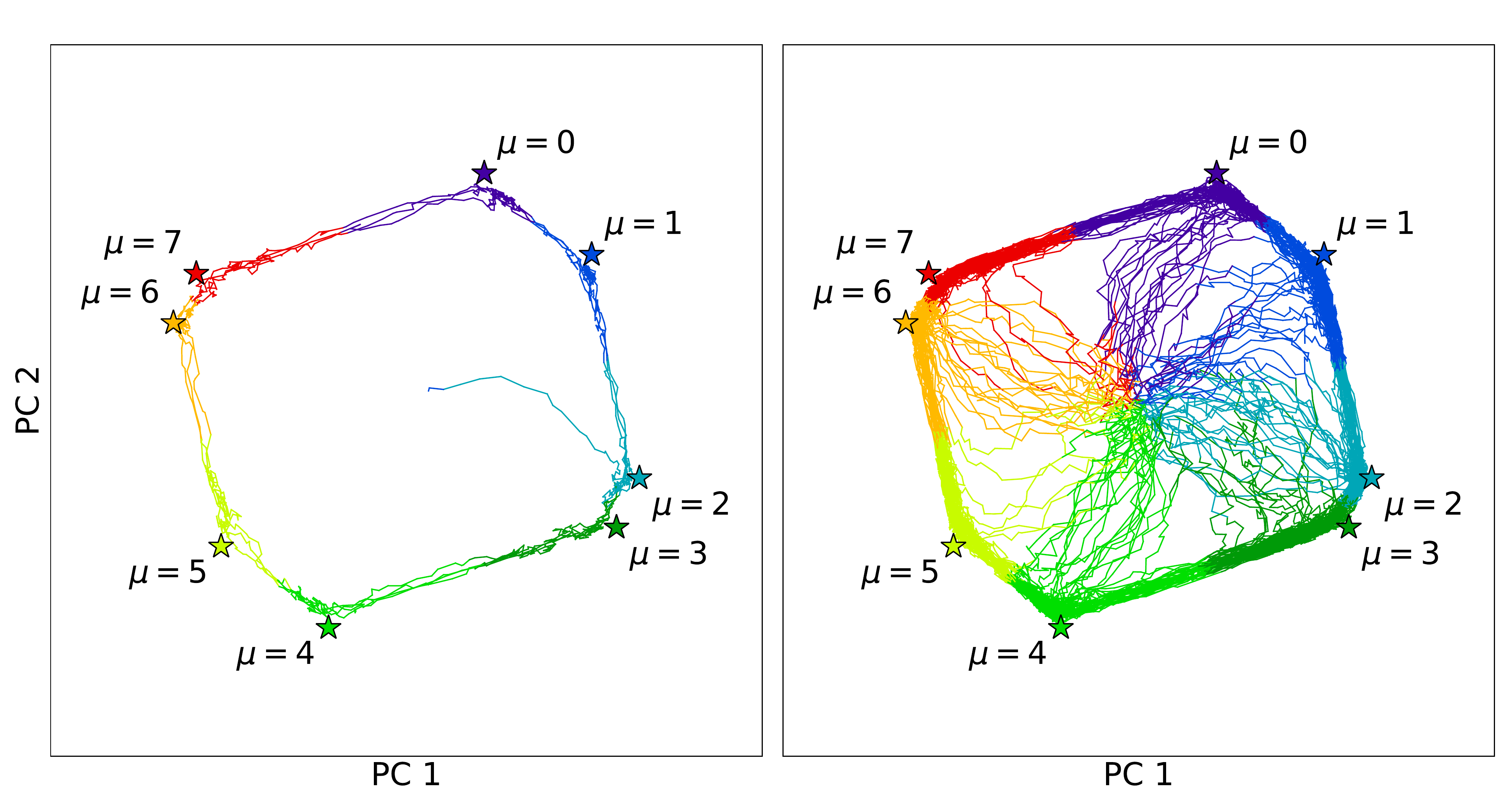}
\end{center}
\caption{{\bf Principal component projection of unperturbed cell trajectories.} The simulated single cell (left panel) and 100 cells (right panel) began with random initial states (projected near the center of the plot), but quickly settled into the encoded cycle. Line segments were colored according to $s_k(t)$, i.e.~which of the eight attractors (labeled stars) had maximum overlap at time $t$.}
\label{fig:exampleTrajectory}
\end{figure}

Trajectories can be visualized by projecting them onto the first two principal components (PCs) of the attractor configurations. Figure~\ref{fig:exampleTrajectory} shows the eight attractors as stars, and a single cell trajectory (left panel) and 100 cell trajectories (right panel) with random initial states as curves with line segments colored according to $s_k(t)$ (as computed in the full $N$-dimensional space). Although the cells begin nearly equidistant from all $\xi_i^\mu$, they quickly relax into encoded cycle. \nameref{sup:random_initial_PCA} shows an animation of a system of $\kappa=50$ cells with random initial conditions projected onto the same PCs, where cells (circles) are colored according to $s_k(t)$. As with the cells shown in Figure~\ref{fig:exampleTrajectory}, all initially random configurations eventually converge to the cycle. \nameref{sup:synch_initial_PCA} shows an animation of $\kappa=50$ cell trajectories with $\sigma_{ik}(0)=\xi_i^0$. As time progresses, the phases of the initially synchronized cells slowly decohere because cells stochastically and independently transition between attractors due to the finite temperature in Eq.~\ref{eq:hopfieldUpdateRule}.

\nameref{sup:temperature_ramp_PCA} demonstrates the effect of temperature on the dynamics. 50 cells were given random initial states, and the temperature was increased and decreased in steps. Cells rarely escape the eight attractor states for $T=0.045$, and one cell becomes stuck near the center in a spurious attractor (unintentional metastable states that arise from the model's nonlinearity). At $T=0.06$, fluctuations allow the cells to transition somewhat regularly through the encoded cycle, and the cell trapped in the spurious attractor eventually escapes and joins the cycle. At $T=0.09$ the cells begin to noticeably diverge from the eight attractor states, but still collectively display a net clockwise flow. The noise is too great for the cells to follow the cycle for $T=0.15$, but lowering the temperature again returns cells to the cycle. This illustrates the fact that the cycle is preserved only for intermediate temperatures: cells become ``frozen'' in intended or spurious attractor states at low temperatures, but at high temperatures the noise is too great and the couplings between genes become irrelevant to the dynamics.

\subsection*{Optimal control fields}

The GA was used to identify some effective combinations of gene targets that slowed progress through the cyclic attractor for varying numbers of targets, $n_\text{targ}$. Because each gene has one of eight discrete phases, there can be multiple equivalent optimal control sets. Here we present and discuss only some of the optimal sets. Extensive tables of results can be found in \nameref{sup:genetic_alg_results}.

\begin{figure}
\begin{center}
\includegraphics[width=1.0\textwidth]{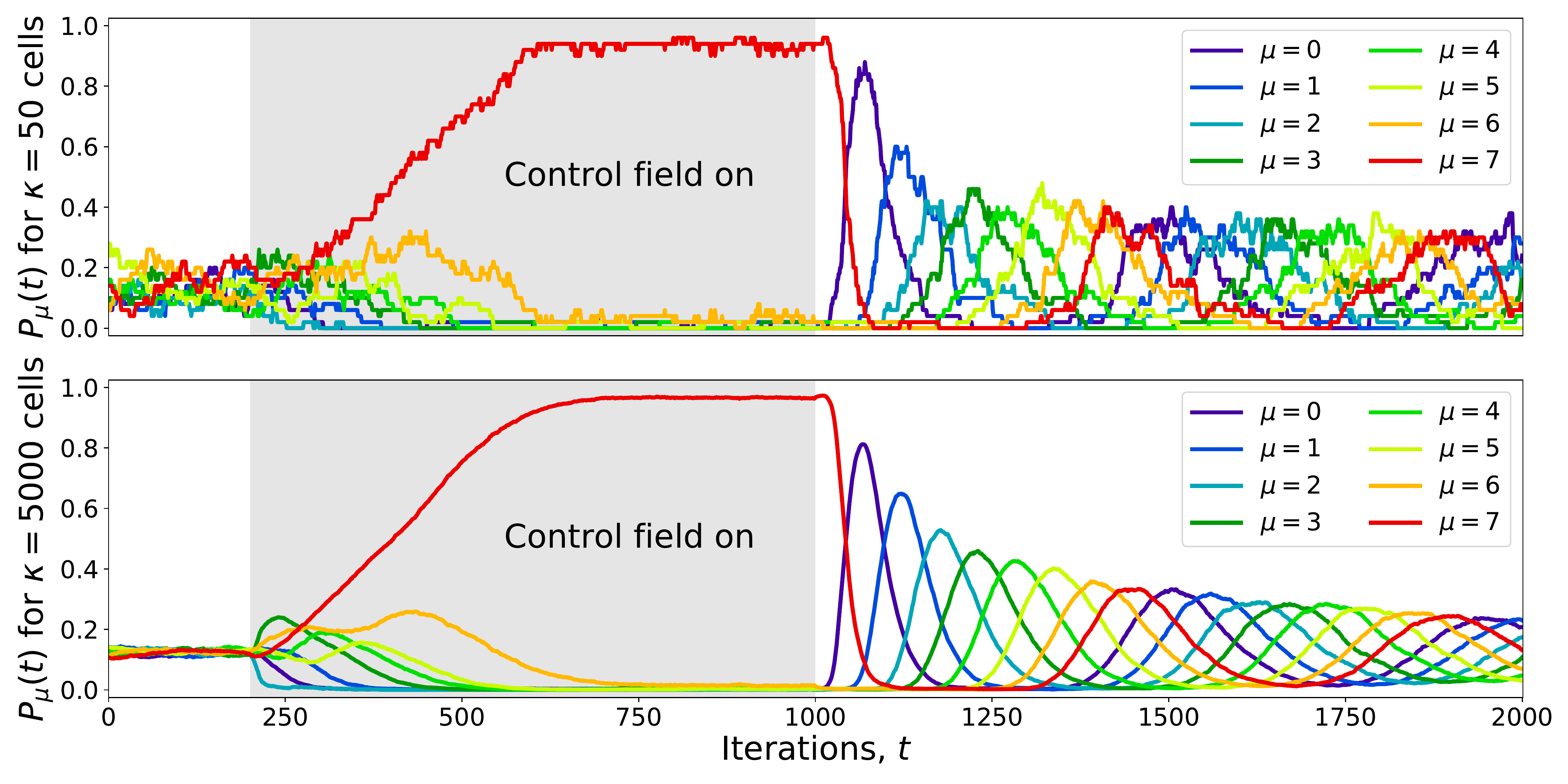}
\end{center}
\caption{{\bf Cell state synchronization by targeted inhibition for 50 and 5000 cells.} Cells were initialized with random states at $t=-200$ (not shown) and allowed to relax into the cyclic attractor so that $P_\mu(0\leq t<200)\approx 1/8$. A set of eight genes was inhibited with an external control field over the range $200\leq t<1000$, fixing most cells near the $\mu=7$ state. After removing the control field, the cells resumed moving through the cycle with initially synchronized phases that slowly broaden. Eventually the system returns to a desynchronized state, $P_\mu(t\rightarrow\infty)\approx 1/8$.}
\label{fig:attractorDistributions}
\end{figure}

The GA found that inhibiting the set of eight \textit{S. cerevisiae} genes \textit{HEK2}, \textit{PRR1}, \textit{QRI1}, \textit{RFC4}, \textit{STB1}, \textit{TDA7}, \textit{VPS17}, and \textit{ZIM17} was sufficient to trap ${\sim}95\%$ of cells in the $\mu=7$ state. The effects of this control field on the time evolution of $P_\mu(t)$ for $\kappa=50$ and $\kappa=5000$ are shown in Figure~\ref{fig:attractorDistributions}. Cells were given random, independent initial states at $t=-200$ (not shown), quickly settling into the cyclic attractor with evenly distributed phases so that $P_\mu(0\leq t<200)\approx1/8$. The control field was activated at $t=200$, causing the cells to accumulate in the $\mu=7$ state. The field was then disabled at $t=1000$, allowing the cells to resume cycling with initially synchronized phases, as shown by the sequence of oscillations in $P_\mu(t>1000)$. The stochastic nature of the transitions causes the cells' phases to slowly spread so that $P_\mu(t\rightarrow\infty)\approx1/8$, eventually returning the system to a desynchronized state.

\begin{figure}
\begin{center}
\includegraphics[width=1.0\textwidth]{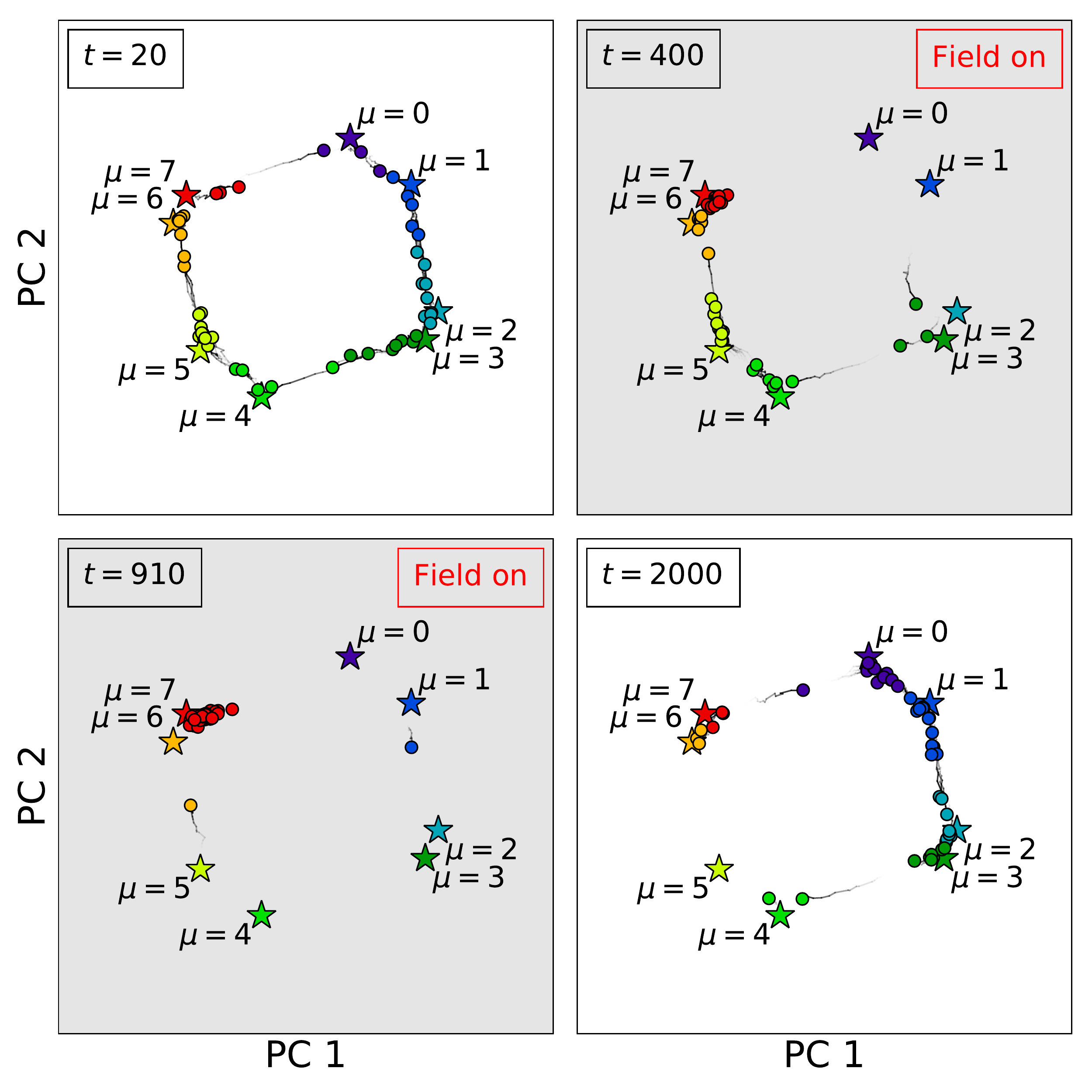}
\end{center}
\caption{{\bf Principal component projection of 50 cell trajectories.} The trajectories used to make the $\kappa=50$ panel of Figure~\ref{fig:attractorDistributions} were projected onto the the first two principal components (PCs) of the attractor array $\xi_i^\mu$ (labeled stars). Cells (circles) are colored according to the closest attractor as computed by Eq.~\ref{eq:cell_state}. When the external field is activated, most cells become trapped in the $\mu=7$ state, although occasionally cells break from the group and complete another circuit before becoming trapped again. After the external field is removed, the cells eventually return to a desynchronized state. See~\nameref{sup:synchronizedPCA} for an animation of these trajectories.}
\label{fig:targetsScreenshots}
\end{figure}

The effects of this control field can also be visualized using a PC projection as shown in Figure~\ref{fig:targetsScreenshots} and~\nameref{sup:synchronizedPCA}. The same set of $\kappa=50$ trajectories from Figure~\ref{fig:attractorDistributions} was projected onto the attractors' PCs, with cells colored according to $s_k(t)$. The control field manages to fix most cells near the $\mu=7$ state, but as shown in the $t=910$ panel in Figure~\ref{fig:targetsScreenshots}, fluctuations occasionally push individual cells out of the $\mu=7$ basin and back into the cycle.

Four further searches were constrained to inhibiting 2, 4, 6, and 8 out of the 24 periodic kinases~\cite{yang2008phosphopoint,hornbeck2011phosphositeplus} in HeLa cells. In all cases, the GA found $\mu=2$ (M phase) to be the most controllable attractor, with an optimal score of $\left\langle P_{2} \right\rangle_\mathscr{T}\approx0.264$ for $n_\text{targ}=4$ (0.8\% of the 519 periodic genes) when inhibiting the genes \textit{BRD4}, \textit{MAPK1}, \textit{NEK7}, and \textit{YES1}; and an optimal score of $\left\langle P_{2} \right\rangle_\mathscr{T}\approx0.281$ for $n_\text{targ}=8$ when inhibiting the genes \textit{BRD4}, \textit{MAPK1}, \textit{NEK7}, \textit{YES1}, \textit{CDC42BPB}, \textit{PRKDC}, \textit{PTK2}, and \textit{TRIM28}. Table~\ref{table:GA_results} shows the results for $n_\text{targ}=4$ for all attractors.

\begin{table}[]
\centering
\resizebox{\textwidth}{!}{%
\begin{tabular}{|c|c|c|c|c|c|c|}
\hline
\textbf{Attr.~index} & \textbf{Approx.~CC phase} & \multicolumn{4}{c|}{\textbf{Kinase inhibition combination}} & \textbf{Score} \\ \hline
0 & G2/M & \textit{MINK1} & \textit{NRBP2} & \textit{OBSCN} & \textit{TAF1} & 0.192 \\ \hline
1 & M & \textit{HIPK1} & \textit{NEK7} & \textit{TAF1} & \textit{YES1} & 0.190 \\ \hline
2 & M & \textit{BRD4} & \textit{MAPK1} & \textit{NEK7} & \textit{YES1} & 0.264 \\ \hline
3 & M/G1* & \textit{CDC42BPB} & \textit{EPHB2} & \textit{MAPK1} & \textit{PTK2} & 0.201 \\ \hline
4 & M/G1* & \textit{EPHB2} & \textit{MNAT1} & \textit{PRKDC} & \textit{PTK2} & 0.197 \\ \hline
5 & G1/S & \textit{CSNK1D} & \textit{HIPK1} & \textit{PANK2} & \textit{PRKDC} & 0.203 \\ \hline
6 & S & \textit{CDK10} & \textit{CDK9} & \textit{CSNK1D} & \textit{MAP4K2} & 0.178 \\ \hline
7 & S/G2* & \textit{CDK10} & \textit{CDK5} & \textit{MAP4K2} & \textit{MINK1} & 0.229 \\ \hline
\end{tabular}%
}
\caption{{\bf HeLa kinase inhibition search results.} The genetic algorithm was used to identify the best combinations of four kinases to inhibit in order to freeze cells in each attractor, along with the optimal score (the time-averaged fraction of cells occupying that attractor state). Phases marked with an asterisk (*) were not found to be significantly enriched for any CC phase, and so are labeled as being between the previous and next known phases.}
\label{table:GA_results}
\end{table}

The structure of the coupling matrix was probed using centrality measures from complex network theory~\cite{newman2010networks} by taking the absolute value of the coupling matrix's elements as edge weights. Katz centrality and PageRank were found to be poor predictors of optimal target sets, but betweenness centrality proved to be a very effective predictor. \textit{BRD4}, \textit{MAPK1}, \textit{NEK7}, and \textit{YES1} have the four highest betweenness centralities in the network (with a mean betweenness centrality of $2.4\times10^{-3}$ and a $p$-value of $4.7\times10^{-4}$, using the inverse of the absolute value of the coupling strengths as weights), indicating that this set of kinases forms a kind of bottleneck in the transmission of signals through the network. Structural network measures, however, do not account for the time-dependent expression of targeted genes, nor how downstream gene expression reacts to upstream perturbations. Controlling nonlinear dynamical systems requires investigating both the structure of the underlying network and the specific form of interactions as defined through the signaling rules.



\section*{Conclusions}

Above we presented a delayed cyclic Hopfield model designed to store CC time series gene expression data from synchronized \textit{S. cerevisiae} and HeLa cells, and the behaviors of both individual cells and populations of cells were studied. The dynamics of populations of cells successfully recreated the experimental gene expression data, including the slow decoherence of initially synchronized cells due to the stochastic transitions between attractors. Optimal control fields that freeze or stall the cyclic attractor by inhibiting only a small number of genes were identified. These predictions could be experimentally validated or invalidated using kinase inhibitors or knockout studies.

Admittedly, there are several limitations to this model. The specific results reported here depend to some degree on the free parameters $T$, $\lambda$, $p$, and the node update probability. Tuning the system to match the behavior of the underlying data places some constraints on these parameters, but a more detailed study of the sensitivity of the results to these parameters could prove useful. Additionally, although using the temporal ordering of the time series gene expression samples provides more information about potentially causal relationships than static samples, the Hopfield model is ultimately an effective model that builds gene-gene couplings from pairwise correlations in gene expression, thereby capturing direct, indirect, and spurious relationships between genes. Independently derived network information with experimentally confirmed molecular regulatory interactions could perhaps be used to refine the construction of the coupling matrix.

Our approach can be generalized and improved in many ways. This incarnation of the model causes simulated cells to continuously undergo CC with no G0 (resting) phase. Adding a relatively stable G0 attractor between the M and G1 phases could cause cells to pause between cycles. A GA search could then be conducted to find the best sets of inhibition targets to freeze cells in the G0 state, or to find the best sets of targets to stimulate entry into CC, mimicking the effects of environmental signals such as growth factors. 

We chose to discretize the continuous gene expression data using a traditional two-state model, which assumes that each gene is either fully activated or fully deactivated. Using a multi-level Hopfield model~\cite{zurada1996generalized} could better reflect the continuous nature of gene expression data and potentially improve the search results. This model can also incorporate additional omics information, e.g.~proteomics and metabolomics, simply by increasing the number of nodes in the system. We plan to explore this option as more multi-omics time series data sets become available. Single-cell experimental techniques and analytical tools are also rapidly improving in quality, decreasing in cost, and gaining in popularity~\cite{patel2014single,scialdone2015computational,zheng2017massively}, and using techniques like pseudo-temporal ordering~\cite{trapnell2014pseudo} could allow the Hopfield model to encode single-cell RNA-seq data as well.

Although the above simulated populations of cells exhibit intriguing dynamical and statistical properties, they behave as completely homogeneous, non-interacting particles. The importance of cell-cell communication and interactions in populations of cells has been demonstrated in a variety of systems including bacterial quorum sensing~\cite{waters2005quorum} and community spatial patterning~\cite{momeni2013strong}, neuron synchronization in circadian rhythm~\cite{kalsbeek2012suprachiasmatic}, and various forms of cancer~\cite{skog2008glioblastoma,hong2009colorectal,renzulli2010microvesicle,del2011marrow,tetta2013extracellular}. As with many nonlinear systems, even seemingly minor changes can produce dramatically different outcomes. More complex extensions to our model could incorporate cell-cell communication by, for example, adding couplings between known signaling molecules and receptors between different cells, and could even allow for interactions between heterogeneous cell types. This would increase the computational complexity of the model, but could better reflect the underlying biology.

\section*{Methods}

\subsection*{Gene expression fitting}

In order to encode these CC data sets into the Hopfield model, periodic genes needed to be identified, their frequencies and phases computed, and their expression converted from continuous to Boolean form. SciPy's Trust Region Reflective method~\cite{scipyreference} was used to identify genes $i$ with periodic expression $x_i(t)$ by fitting to the form
\begin{equation}
x_i(t) = a_ie^{-b_it}\cos\left(\omega_i t - \phi_i\right) + x_{i0}
\label{eq:firstFit}
\end{equation}

\noindent for amplitude $a_i$, decay rate $b_i$, angular frequency $\omega_i$, phase $\phi_i$, and asymptotic mean expression $x_{i0}$. (Because the HeLa data set has fewer time points (14) than the \textit{S. cerevisiae} data set (41), analysis of the HeLa data set was preceded by a smoothing step using a simple box filter to aid in fitting.) The first several time points were ignored to avoid fitting the parameters of Eq.~\ref{eq:firstFit} to chemically perturbed (transient) states. A gene was labeled periodic if the maximum relative uncertainty in its parameters from the fit,
\begin{equation}
r_i^\text{max} = \max \left\{
\frac{\delta x_{i0}}{x_{i0}},
\frac{\delta a_i}{a_i},
\frac{\delta b_i}{b_i},
\frac{\delta \omega_i}{\omega_i},
\frac{\delta \phi_i}{2\pi}
\right\} \text{ ,}
\end{equation}

\noindent was less than the thresholds defined in~\nameref{S1_Table}. Once all frequencies $\{\omega_i\}$ for periodic genes were computed, the frequency was fixed to the mean frequency $\bar{\omega}$ and the fits were recomputed for each periodic gene using the form
\begin{equation}
x_i(t) = a_ie^{-b_it}\cos\left(\bar{\omega} t - \phi_i\right) + x_{i0} \text{ ,}
\label{eq:fitWithFixedFrequency}
\end{equation}

\noindent thus producing the final set of continuous phases $\{\phi_i\}$. Figure~\ref{fig:snakePlot}A shows a heat map of the expression of all periodic genes detected in the Eser data set sorted by their fitted phases, and Figure~\ref{fig:snakePlot}B shows the same genes with the fitted expression curves. These fitted curves were converted from continuous values $x_i(t)\geq0$ to Boolean values $\xi_i(t)=\pm1$ (over/underexpressed) by assigning 
\begin{equation}
\xi_i(t) = \sign(x_i(t)-x_{i0})
\end{equation}

\noindent as shown in Figure~\ref{fig:snakePlot}C. Finally, one CC period was divided into eight uniformly spaced states $\{\xi_i^\mu\} = \{\xi_i^0,\xi_i^1,\ldots,\xi_i^7\}$. These states, shown in Figure~\ref{fig:snakePlot}D, were used as attractors in the Hopfield model.

\subsection*{Determining cell cycle phase}

To determine the approximate CC phases for each attractor $\mu$ in the HeLa data set, over-representation analysis was conducted using the hypergeometric distribution to calculate $p$-values with the Benjamini-Hochberg procedure~\cite{hochberg1990more} to correct for multiple hypothesis testing with false discovery rate $<0.05$, using all genes $i$ with $\xi_i^{\mu-1}=-1$ and $\xi_i^\mu=+1$ as the $\mu^\text{th}$ input set and using all detected cyclic genes as the background. In the Dominguez HeLa data set, $\mu=0$ was enriched for the gene ontology (GO) terms ``negative regulation of cell proliferation'' ($p=\num{2.3e-2}$) and ``DNA double-strand break repair'' ($p=\num{3.1e-2}$), corresponding to the G2/M checkpoint. $\mu=2$ was enriched for the GO term ``nuclear envelope breakdown'' ($p=\num{4.2e-3}$), corresponding to the preparation of chromosome condensation and cellular mitosis. $\mu=5$ (``\textit{TP53} regulates transcription of DNA repair genes,'' $p=\num{5.0e-3}$) and $\mu=6$ (``DNA strand elongation,'' $p=\num{2.0e-3}$) correspond to the G1/S phase checkpoints and the elongation of DNA in S-phase respectively. The database \url{yeastgenome.org}~\cite{cherry2011saccharomyces} was used to determine the CC phases for the Eser \textit{S. cerevisiae} data set. $\mu=0$ is enriched for the GO term ``DNA replication'' ($p=\num{2.02e-12}$), indicating an attractor in the S phase of CC. $\mu=2$ is enriched for ``mitotic spindle organization'' ($p=\num{2.3e-3}$) indicating the beginning of mitosis in \textit{S. cerevisiae}. $\mu=6$ from Eser is enriched for the GO term ``pre-replicative complex assembly involved in nuclear cell cycle DNA replication'' ($p=\num{3.0e-5}$), indicating an attractor at the end of G1 phase as the cells prepare for DNA replication.

\section*{Supporting Information}





\subsection*{S1 Video}
\label{sup:random_initial_PCA}
{\bf 50 cell trajectories with random initial conditions.} Data was projected onto the first two principle components of the attractor array $\xi_i^\mu$. Attractors are shown as stars, and cells are shown as circles. Cell colors are assigned using $s_k(t)$ as measured in the full $N$-dimensional space. All cells $k$ were given random initial conditions $\sigma_{ik}=\pm1$ with equal probability for all $i$ and $k$, but eventually converge to the cyclic attractor.


\subsection*{S2 Video}
\label{sup:synch_initial_PCA}
{\bf 50 cell trajectories with identical initial conditions.} See the caption of~\nameref{sup:random_initial_PCA} for an explanation of the projection and colors. All cells $k$ were initially synchronized with $\sigma_{ik}(t)=\xi_i^0$, but progress through the cycle stochastically, causing the distribution of $s_k(t)$ to broaden.


\subsection*{S3 Video}
\label{sup:temperature_ramp_PCA}
{\bf Effects of temperature on 50 cell trajectories.} See the caption of~\nameref{sup:random_initial_PCA} for an explanation of the projection and colors. All cells were given initial random states, and the temperature was increased and decreased in steps as shown in the top panel.


\subsection*{S4 Video}
\label{sup:synchronizedPCA}
{\bf Principal component projection of 50 cells being synchronized.} See the caption of~\nameref{sup:random_initial_PCA} for an explanation of the projection and colors. This video is an animation of the trajectories used in Figures~\ref{fig:attractorDistributions} and~\ref{fig:targetsScreenshots}.



\subsection*{S1 Text}
\label{sup:genetic_alg}
{\bf Explanation of genetic algorithm.}

\subsection*{S1 Table}
\label{S1_Table}
{\bf List of parameters used for each data set.}

\subsection*{S1 File}
\label{sup:genetic_alg_results}
{\bf Zip file containing results of genetic algorithm's gene inhibition searches.}


\section*{Acknowledgments} 

We would like to thank the National Institutes of Health for supporting this work through the grant NIH/NIGMS R01GM122085. We would also like to thank George Mias (Michigan State University) and Yunyi Kang (Sanford Burnham Prebys Medical Discovery Institute) for helpful discussions and recommendations.


\bibliography{BIB}{}


%
%

\end{document}